\documentclass[a4paper,11pt]{article}
\usepackage{aaskaiid}
\usepackage{orcidlink}

\title{Unraveling the mysteries of supernovae with SKA+VLBI}
\ShortTitle{Supernovae with SKA+VLBI}

\author[1,2\ast]{Tao An \orcidlink{0000-0003-4341-0029}}
\ShortName{An et al.} 
\author[3,\ast]{Zhengwei Liu \orcidlink{0000-0002-7909-4171} }
\author[3,4,5]{Ailing Wang \orcidlink{0000-0002-7351-5801} } 
\author[6]{Miguel Pérez-Torres \orcidlink{0000-0001-5654-0266} }
\author[6]{Javier Moldon \orcidlink{0000-0002-8079-7608}}
\author[7]{Liangduan Liu \orcidlink{0000-0002-8708-0597}}
\author[8]{Peter Lundqvist \orcidlink{0000-0002-3664-8082}}
\author[3]{Xuefei Chen \orcidlink{0000-0001-5284-8001}}
\author[3]{Xiangcun Meng \orcidlink{0000-0001-5316-2298}}

\affiliation[1]{Department of Astronomy, University of Science and Technology of P.~R.~China, Hefei, Anhui 230026, P.~R.~China}
\affiliation[2]{Shanghai Astronomical Observatory, Chinese Academy of Sciences, Shanghai 200030, P.~R.~China}
\emailAdd{antao2008@ustc.edu.cn}
\affiliation[3]{International Centre of Supernovae (ICESUN), Yunnan  Key Laboratory of Supernova Research, Yunnan Observatories, CAS, Kunming 650216, P.~R.~China}
\emailAdd{zwliu@ynao.ac.cn}
\affiliation[4]{State Key Laboratory of Particle Astrophysics, Institute of High Energy Physics, Chinese Academy of Sciences, Beijing 100049, P.~R.~China}
\affiliation[5]{Spallation Neutron Source Science Center, 523803 Dongguan, P.~R.~China}
\affiliation[6]{Instituto de Astrof\'isica de Andaluc\'ia (IAA-CSIC), Glorieta de la Astronom\'ia s/n, E-18008 Granada, Spain
}
\affiliation[7]{Institute of Astrophysics, Central P.~R.~China Normal University, Wuhan 430079, P.~R.~China}
\affiliation[8]{The Oskar Klein Centre, Department of Astronomy, Stockholm University, AlbaNova, SE-10691 Stockholm, Sweden}
\affiliation[\ast]{Chapter Corresponding Author}

\abstract{
Supernovae (SNe) drive cosmic chemical enrichment and shape galactic feedback, yet the link between  progenitors and explosion outcomes remains poorly constrained because the earliest phases are rarely resolved. Radio emission traces synchrotron radiation where the fastest ejecta interact with the circumstellar medium (CSM), providing a uniquely penetrating probe of these phases. SKA‑Mid phased into global VLBI will move from simple detections to routine interferometric imaging of nearby extragalactic SNe. Sub-$\mu$Jy sensitivity and  mas-scale SKA+VLBI imaging, complemented by visibility-domain model fitting for sub-beam radius measurements at 5--15~GHz will allow us to follow the expanding shocks of stripped-envelope SNe out to $\sim$25 Mpc, measure deceleration indices ($m$) and axial ratios to $\approx 5-10\%$, and directly test jet assisted versus neutrino driven explosion mechanisms. For interacting SNe (Type IIn/Ibn), SKA+VLBI will resolve clumpy and toroidal CSM on progenitor scales, constraining the timing and geometry of eruptive pre explosion mass loss. Deep limits on Type Ia SNe will tightly restrict the allowed single-degenerate parameter space, while late-time imaging will search for nascent compact remnants and pulsar wind nebulae. In synergy with optical, X-ray and gravitational wave facilities, SKA+VLBI will turn nearby SNe into laboratories for time resolved shock physics and progenitor mapping.
}

\begin{document}
\maketitle

\section{Introduction \& Context}
Supernovae (SNe) are among the most energetic events in the Universe \citep{1996snih.book.....A, 2005NatPh...1..147W}. They end stellar lives with kinetic energy of order $\sim10^{51}$ erg, enrich the interstellar medium (ISM) with heavy elements, and launch shocks that seed and regulate cosmic‑ray populations. Depending on the physical explosion mechanisms, progenitor channels and environments, SNe fall into three main categories: thermonuclear supernovae, i.e. thermonuclear explosions of white dwarfs (Type Ia supernovae, SNe Ia), core-collapse supernovae (CCSNe) and superluminous supernovae (SLSNe).  
Using SNe Ia as accurate cosmic distance indicators led to the discovery of the accelerating expansion of the Universe \citep{1998AJ....116.1009R, 1999ApJ...517..565P}, winning the Nobel Prize in Physics for 2011. Despite their importance as cosmic engines and standard candles, fundamental questions remain about the nature of progenitor systems and the explosion physics across different subclasses \citep{2012ARNPS..62..407J, 2013RvMP...85..245B, 2014ARA&A..52..107M, 2012Sci...337..927G}. Radio observations are uniquely poised to close these gaps by directly tracing how the fastest ejecta interact with their environments in the time domain \citep{ 2002ARA&A..40..387W, 2006ApJ...651..381C, Chandra01.2026.SKA}. 

When a supernova’s rapidly expanding ejecta plough into the circumstellar medium (CSM) and ISM, electrons are accelerated and magnetic fields amplified, producing long‑lived synchrotron emission that is robust against dust extinction \citep{2002ARA&A..40..387W, 2006ApJ...651..381C}. This radio glow encodes the properties of the progenitor’s  CSM and/or ISM (e.g. stellar winds) and the structure of the explosion’s shock \citep[e.g.,][]{2002ARA&A..40..387W,2006ApJ...651..381C}, including its degree of asymmetry and time-dependent evolution. Therefore, radio emission provides one of the most sensitive and efficient probes of the external density, shock microphysics, and geometry, placing stringent constraints on the origins of different types of SNe from ordinary core-collapse explosions to rare SLSNe and interacting Ia–CSM events.

The basic theoretical framework is well established: synchrotron self--absorption (SSA) governs the early spectral turnover \citep{1982ApJ...258..790C, 1998ApJ...499..810C}. For steady winds, the CSM density scales as $\rho_{\rm CSM}(r)\propto r^{-2}$; deviations (such as rebrightenings, plateaus, sharp spectral breaks) betray clumps, equatorial enhancements, or eruptive shells \citep{2011ApJ...729L...6C, 2012ApJ...752L...2C, 2014ARA&A..52..487S}. Very Long Baseline Interferometry (VLBI) provides significant advantages by resolving structures on milliarcsecond (mas) scales. Limb-brightened rings, azimuthal brightness modulations and compact central components can be \textit{measured}, rather than inferred. Historically, however, sensitivity and angular resolution limited radio SNe to a few dozen detections and only a handful of resolved shells. Iconic VLBI campaigns, SN~1993J most notably, have imaged the expanding shell and its kinematics over years, constraining deceleration and revealing azimuthal brightness modulations indicative of asymmetry and CSM structure \citep{2001ApJ...557..770B, 2003ApJ...597..374B, 2009A&A...505..927M}. Subsequent VLBI on young IIb events such as SN~2011dh extended this to the first few hundred days \citep{2011A&A...535L..10M, 2012ApJ...751..125B, 2016MNRAS.455..511D}. For Type~Ia, deep non--detections have imposed stringent mass-loss limits on the bulk population \citep[e.g.,][]{2006ApJ...646..369P, 2012ApJ...750..164C, 2014ApJ...792...38P}, while the radio-detected Ia--CSM event SN~2020eyj highlights a rare helium-rich channel that demands population-level context \citep{2023Natur.617..477K}. These facts motivate a facility that can both \textit{find} faint radio SNe promptly and \textit{resolve} their shocks routinely for a much larger and more diverse SN population.

SKA--Mid, coupled to global VLBI networks, provides precisely this capability by combining wideband survey speed with phased-array sensitivity on intercontinental baselines. SKA--Mid spans $\approx0.35$--$15.4$ GHz; hour--scale continuum imaging achieves rms sensitivities at the few $\mu$Jy beam$^{-1}$ level across Bands~2--5 \citep{SKAO_perf}, enabling rapid spectral follow--up from SSA rise to optically thin decline for nearby SNe throughout the local 30--50 Mpc volume. In phased--array mode, SKA's collecting area is tied to intercontinental baselines, yielding synthesized beams of sub-mas to mas (typically $\approx5$ mas at 1.6 GHz, $\approx1.5$ mas at 5 GHz, $\approx0.9$ mas at 8 GHz and $\approx0.6$ mas at 15 GHz on global baselines). At 10 Mpc, $1$ mas $\approx 0.05$ pc, so sub-parsec shocks are resolvable within $\lesssim20$ Mpc, while model-fitting in the visibility domain recovers radii below the nominal beam given $S/N\gtrsim10$. Throughout this chapter, `imaging' refers to measurements at the synthesized-beam scale, whereas `sub-beam' size constraints refer to visibility-domain model fitting (super-resolution) enabled by high S/N. In parallel, wide-field time--domain surveys (ZTF today; Vera C.\ Rubin Observatory's LSST imminently) supply explosion times and early classifications \citep{2019PASP..131a8002B,2019ApJ...873..111I}, allowing SKA to trigger within 24~h and to schedule multi-epoch VLBI that captures both kinematics (expansion, deceleration) and morphology (axial ratio, position angle stability). Together, these capabilities transform radio SNe from point sources into spatially resolved laboratories, where geometry, microphysics, and progenitor mass-loss histories can be measured with precision across CCSNe, the rare Ia--CSM channel, and the debated central-engine scenarios for SLSNe in a fully time-resolved, multi-messenger context.

\section{Supernovae with SKA+VLBI}
\label{sec:sne_vlbi}

\subsection{Motivation and unique capability}

VLBI adds the missing spatial dimension: it directly images the synchrotron-emitting forward shock, converting radio light curves into measurements of shock radius, deceleration, and morphology that diagnose both explosion geometry and circumstellar structure. 
Until now, the field has been limited primarily by sensitivity and angular resolution: only a few dozen SNe have been detected in the radio band, and only a small subset (for example SN 1993J, SN 2011dh) have been resolved with VLBI.  
SKA--Mid phased into the global VLBI network removes this bottleneck and turns nearby SNe from unresolved transients into directly imaged blast waves.  

At 8 GHz, a synthesized beam of $\approx1$ mas corresponds to $\simeq0.05$ pc at 10 Mpc, and model fitting in the visibility domain recovers even smaller effective radii when ${\rm S/N}>10$.  For clarity, we quote two relevant angular scales: the synthesized beam (true imaging resolution) and the smaller radii obtainable via visibility-domain shell fitting at high S/N \citep{2012A&A...541A.135M, 2014A&A...563A.136M}. 
Hour‐scale integrations yield $\sigma_{\rm rms}\lesssim 2\,\mu$Jy beam$^{-1}$ (Band 5), enabling the detection and imaging of ordinary core‐collapse SNe throughout the local 30--50 Mpc volume.  

A typical shock with velocity $v_{\rm sh}$ expands at an angular rate as
\begin{equation}
    \mu \simeq 0.58~\mu{\rm as\,d^{-1}}
    \left( \frac{v_{\rm sh}}{10^4~{\rm km\,s^{-1}}} \right)
    \left( \frac{D}{10~{\rm Mpc}} \right)^{-1},
\end{equation}
so the angular radius grows by $\sim17~\mu$as in a month at 10 Mpc. This scale is resolvable by SKA+VLBI on timescales of weeks to months, turning \emph{single‐pixel} light curves into time-resolved ``shock cinematography''. Direct radius measurements, in combination with broadband spectra, will allow us to infer energy partition, asymmetry, and the CSM density profile.

\subsection{Explosion geometry and ejecta asymmetry}

\textit{Key questions: how axisymmetric are nearby explosions, and does the degree of asymmetry correlate with subtype or environment?}

Optical polarimetry and nebular‐line profiles already show that stripped‐envelope SNe (IIb/Ib/Ic) are often non‐spherical (\citealt{2008Sci...319.1220M, 2008ARA&A..46..433W}), but these diagnostics probe the inner ejecta rather than the forward shock.  
VLBI uniquely isolates the outer shock front, where hydrodynamic instabilities or jet‐like energy deposition produce azimuthal brightness modulation \citep{2001ApJ...557..770B, 2003ApJ...597..374B, 2009A&A...505..927M, 2011A&A...535L..10M, 2012ApJ...751..125B, 2016MNRAS.455..511D}.  Mapping this layer is essential to test whether a fraction of stripped-envelope SNe are genuinely jet assisted, as suggested by broad-lined Ic events and the low-luminosity GRB population.

Multi‐epoch SKA+VLBI imaging at 5--8 GHz with cadences of 10--60 d for SNe within 10--20 Mpc will measure the expansion law $R(t)\propto t^m$ and the projected axial ratio $a/b$ to $\approx5$--10\% precision.  
A statistical sample will determine whether a large fraction of IIb/Ib/Ic events exhibit $a/b \ge 1.3$ within the first 100 d, supporting jet‐assisted energy deposition, or whether most explosions remain quasi‐spherical with $a/b \lesssim 1.2$.  
Joint analysis with optical polarization and nebular line profiles will then link the outer-shock geometry to the inner ejecta structure, providing a physically grounded bridge between explosion models and observed asymmetries.

Disentangling intrinsic CSM asymmetry from explosion-driven ejecta geometry requires exploiting observables that respond differently to boundary shape and to emissivity around the rim. In practice we will (i) fit parametric thin-shell or ring models directly to the visibilities to recover the shock boundary and its centre independent of azimuthal brightness, (ii) track the stability of axial ratio and position angle with time and frequency, since ejecta-driven bipolarity should produce a coherent, persistent low-order mode while CSM clumps or large-scale density gradients tend to generate evolving hotspots and frequency-dependent absorption, and (iii) search for azimuthal variations in the deceleration parameter and spectral index around the shell. Combined with optical polarimetry and nebular diagnostics, this provides a practical route to separate explosion asymmetry from environmental sculpting.

\subsection{Mass‐loss histories and CSM structure}

\textit{Key questions: what mass-loss histories and angular structures surround SN progenitors, and how do they shape shock microphysics?}

For a steady wind, the CSM density scales as $\rho_{\rm CSM}(r)\propto r^{-2}$ and the synchrotron self‐absorption (SSA) turnover evolves approximately as 
$\nu_p\propto t^{-1}$, $F_p\propto t^{1.2}$ for adiabatic expansion \citep{1998ApJ...499..810C, 2006ApJ...651..381C}.  
The observed triplet $(\nu_p,F_p,t_p)$ then yields an effective wind density parameter:
\begin{equation}
    \frac{\dot{M}}{v_w} \approx 3\times10^{-6}
    \left( \frac{F_p}{1~{\rm mJy}} \right)^{1.3}
    \left( \frac{D}{10~{\rm Mpc}} \right)^{2.6}
    \left( \frac{\nu_p}{5~{\rm GHz}} \right)^{-1.1}
    \left( \frac{t_p}{30~{\rm d}} \right)^{1.3}
    {\rm M_\odot\,yr^{-1}\,(10~km\,s^{-1})^{-1}},
\end{equation}
where $\dot{M}$ is pre-SN mass-loss rate and $v_w$ the wind velocity. The numerical coefficient in Eq.~(2) implicitly assumes fiducial microphysics (electron index and energy fractions); for SSA-dominated solutions the inferred $\dot{M}/v_w$ depends on the magnetic-field energy fraction approximately as $(\dot{M}/v_w)\propto\epsilon_B^{-1}$ (with weaker dependence on $\epsilon_e$), so an independent radius measurement is needed to constrain $\epsilon_B$ and density separately.
Combining these spectral fits with VLBI‐measured radii breaks the degeneracy between magnetic‐field fraction $\epsilon_B$ and density, allowing direct reconstruction of pre‐SN mass‐loss histories over the last $\approx 10-10^3$ yr before core-collapse. 

SKA’s wide frequency coverage and polarization capability will map limb‐brightened shells, clumpy equatorial enhancements, and toroidal geometries in interacting SNe IIn/Ibn within $\lesssim50$ Mpc, revealing eruptive LBV‐like shells months before explosion. Axis-dependent spectra (polar versus equatorial lines of sight) and small-scale brightness substructure will test whether the CSM is predominantly toroidal, clumpy, or multi-shell \citep{2011ApJ...729L...6C}. Targeted polarimetry (when S/N permits) will further constrain post-shock Faraday rotation and magnetic field ordering.
Because young SNe typically exhibit very low intrinsic linear polarization (often $\lesssim1\%$) due to turbulent fields and Faraday depolarization, meaningful polarimetry requires very high Stokes-$I$ S/N. For example, a 5$\sigma$ detection of a 1\% polarized fraction requires ${\rm S/N}\gtrsim500$ in Stokes $I$, so polarimetry will be feasible mainly for the brightest nearby events (sub-mJy to mJy) and/or via longer integrations or stacking of multiple epochs; for fainter tens-of-$\mu$Jy SNe we will primarily obtain stringent upper limits.

On a population level, there is a long-standing ``SNe rate problem'': the core-collapse SN rate inferred from star formation is higher than that measured in optical surveys, particularly at higher redshift  \citep{2011ApJ...738..154H}. The missing fraction of CCSNe increases from $30\%$ to $60\%$ as the redshift increases from $z\approx1$ to $z\approx2$  \citep{2007MNRAS.377.1229M, 2012ApJ...756..111M}, likely because of dust obscuration and crowding in galactic centres. SKA surveys will recover a substantial fraction of these dust-obscured CCSNe in nearby and intermediate-redshift nuclear star-forming galaxies, where individual radio SNe remain detectable, thereby improving the CCSN census and tightening constraints on the cosmic star-formation history (see also \citealt{2015aska.confE..60P}). At $z\gtrsim1$, however, ordinary CCSNe are expected to be too faint for individual detections, and SKA will preferentially detect the most radio-luminous interacting or engine-driven events or constrain the population through stacking.

\subsection{Compact central sources and pulsar birth}

\textit{Key questions: do newborn neutron stars produce detectable radio pulsar wind nebulae (PWNe) within $\lesssim1$ yr, and what are their birth fraction and energetics?}

Newly formed neutron stars may power PWNe within the first year \citep{2006ARA&A..44...17G}, contributing a flat or inverted spectrum and a very compact ($\lesssim 0.2$--$0.3$ mas) component either at the explosion center or modestly offset by natal kicks. Whether such a component is observable within $\lesssim1$ yr depends on free--free absorption through the expanding ejecta and on its ionization state; in many CCSNe the ejecta may remain opaque at GHz frequencies for years. Our emphasis on 8--15 GHz VLBI and on low-ejecta-mass stripped-envelope events targets the most favourable cases where a PWN can emerge within months to a year, while non-detections will still place quantitative limits on early spin-down power and ejecta transparency. A fraction of CCSNe may therefore reveal \textit{compact, flat/inverted-spectrum cores} as nascent PWNe \citet{Gelfand01.2026.SKA}. SKA+VLBI will use source compactness, spectra and variability to distinguish such central engines from shock hotspots along the shell. Monitoring nearby Type IIP and stripped-envelope SNe within $\le20$ Mpc on time scales of $\approx$ 100-300 d post explosion will place the first systematic constraints on the birth fraction of radio-loud pulsars and on their initial spin periods via the PWN luminosity evolution \citep{2006ARA&A..44...17G}. A handful of clean detections would directly calibrate the  initial spin--down power distribution; a null result across a sizeable nearby sample would instead imply delayed emergence  or intrinsically faint or slowly rotating pulsars at birth. Either outcome addresses a long-standing debate about how quickly PWNe become visible.

Operationally, we will search for the appearance of a stationary, high brightness temperature component as the expanding shell becomes optically thin. Simultaneous high-frequency coverage (8--15 GHz) mitigates free-free absorption and interstellar scattering, maximising the detectability of compact central sources and enabling robust constraints on their brightness temperatures.

\subsection{Luminosity sources of superluminous supernovae}

\textit{Key questions: are SLSNe generally powered by magnetars, and if so, when and how do their radio nebulae emerge?}

SLSNe, recognized only a decade ago \citep[e.g.,][]{2009Natur.462..624G, 2012Sci...337..927G}, are an order of magnitude brighter than other ordinary CCSNe and their luminosity source remains not yet clarified \citep[e.g.,][]{2018SSRv..214...59M}. The leading model invokes spin down of strongly magnetized neutron stars (a magnetar, \citealt{Kasen2010}), but there is still no direct evidence for magnetars embedded inside SLSNe. Magnetar models predict that SLSNe can become bright in radio on time scales of $\approx 1$ month to several hundred days, either through an internal PWN-like nebula or through the interaction of magnetar driven relativistic ejecta with the CSM. 
For statistical studies we target SLSNe within $D\lesssim200$ Mpc (roughly $z\lesssim0.05$), where SKA can reach the required $\mu$Jy sensitivities in practical integrations; exceptionally nearby events ($D\lesssim50$ Mpc) are rare but would enable the strongest VLBI constraints on compactness.

To date, deep radio searches have not yielded a robust detection of SLSNe \citep{2018ApJ...856...56C, 2019ApJ...876L..10E}, leaving the magnetar scenario only weakly tested. 
SKA will assemble the first statistical radio sample of SLSNe, probing engine power, baryon loading, and the presence or absence of moderately relativistic outflows.  Deep follow-up beginning $\approx30$ d post-explosion and extending to $\sim200$ d at 8--15 GHz will either reveal the predicted radio-bright phase or decisively constrain it. VLBI-detected source compactness tests whether detections are truly central-engine powered or instead CSM-interaction dominated at larger radii. A yes/no answer at scale transforms the magnetar debate from model-driven to \textit{data--driven}.

\subsection{Progenitors of Type Ia supernovae}

\textit{Key questions: what fraction of SNe~Ia interact with dense CSM (the ``Ia--CSM'' channel), and how stringent can we make the mass-loss limits for the bulk population?}

SNe Ia are widely believed to result from thermonuclear explosions of carbon oxygen white dwarfs (WDs) whose masses approach the Chandrasekhar limit. It is still unclear \citep{2014ARA&A..52..107M} whether the explosion is triggered by mass accretion from a non-degenerate star (the single-degenerate scenario, SDS) \citep[e.g.,][]{2004MNRAS.350.1301H, 2013A&A...552A..24B} or by the merger with two white dwarfs (the double-degenerate scenario, DDS) \citep[e.g.,][]{1973ApJ...186.1007W, 1984ApJS...54..335I, 1984ApJ...277..355W, 2024ApJ...972..200B, 2025arXiv250716757S}. In the SDS, substantial outflows from the donor or the accretion disc are expected to build a dense CSM, whereas in the DDS little or no nearby CSM is expected and prompt radio emission from CSM interaction should be weak or absent. This makes sensitive radio observations a powerful discriminant between the two channels. 

Decades of radio non-detections for most nearby SNe Ia have established stringent upper limits on CSM density and have ruled out many symbiotic and SDS progenitor models with high-$\dot M$, for example in SN 2011fe \citep{2012ApJ...750..164C} and SN 2014J \citep{2014ApJ...792...38P}. That narrative shifted recently with SN 2020eyj, an SNe Ia event interacting with helium-rich CSM, which became the first radio-detected SNe Ia and provide direct evidence that at least one helium--rich Ia CSM channel exists \citep{2023Natur.617..477K}. 

\begin{figure}
    \centering
\includegraphics[width=0.7\linewidth]{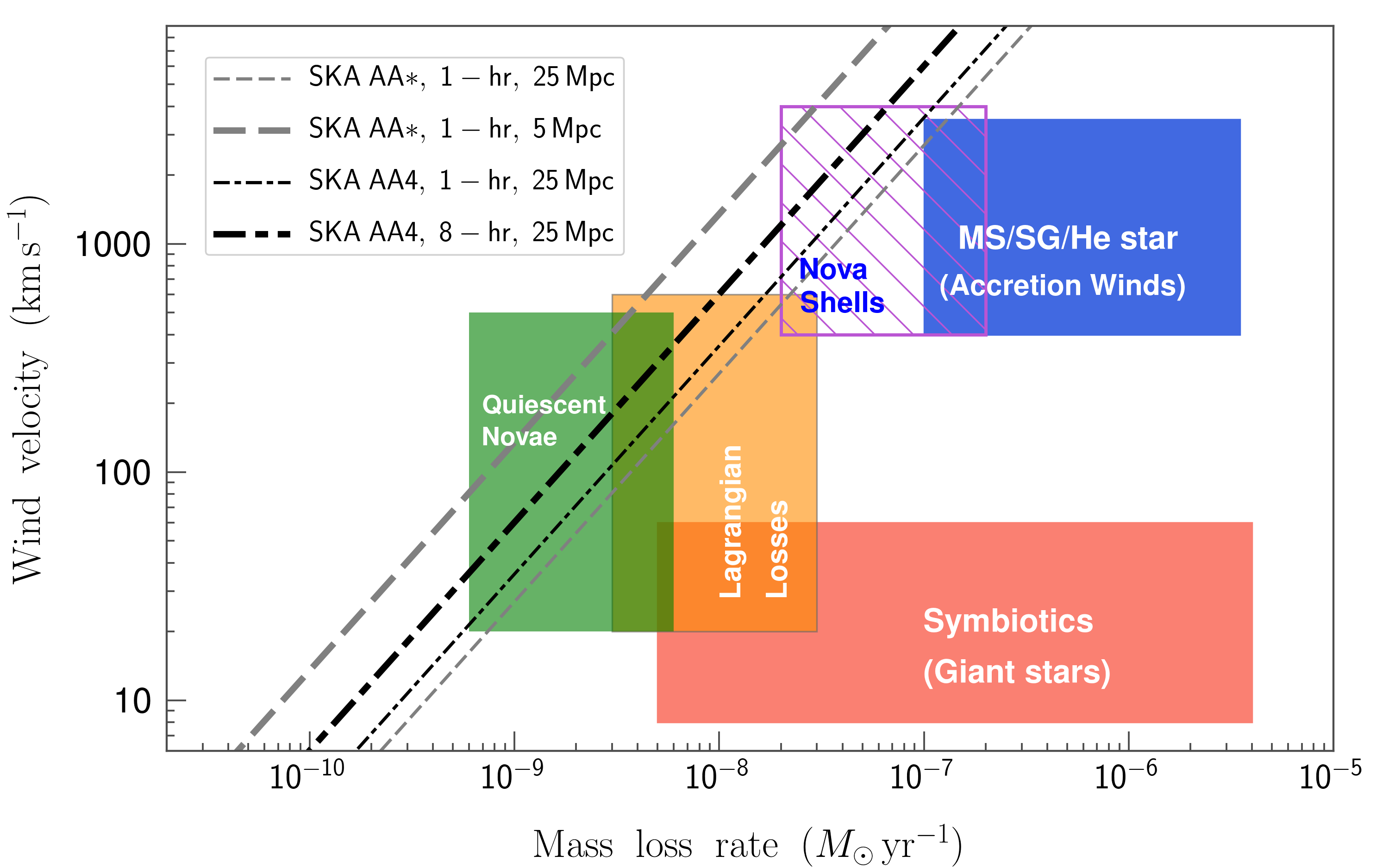}
    \caption{Pre–supernova mass-loss parameter space for single-degenerate SNe~Ia progenitors. The coloured boxes mark the approximate ranges of mass-loss rate and wind velocity expected for quiescent novae, Lagrangian losses in Roche-lobe–overflow systems, symbiotic binaries with giant donors, and main-sequence / subgiant / He-star donors with accretion-driven winds; the hatched magenta region indicates nova shells. Diagonal lines show representative 3$\sigma$ detectability thresholds for the indicated integration times and distances, assuming standard microphysics. The boundaries are approximately lines of constant $\dot{M}/v_w$, reflecting that radio emission primarily depends on the wind density parameter; thus higher $v_w$ requires proportionally higher $\dot{M}$ for detectability. Systems lying above a given line would yield detectable radio emission in that configuration, so deep non-detections with SKA+VLBI can exclude large parts of the single-degenerate parameter space and pinpoint rare Ia--CSM channels.}
    \label{fig1}
\end{figure}

SKA will extend this frontier. Rapid high-sensitivity (a few $\mu$Jy) follow-up within 5--15 days will clearly separate rare radio detections (constraining the Ia--CSM fraction) from deep non-detections that further tighten mass-loss limits for the bulk SNe Ia population. With 5--15 d triggers and 5--15 GHz coverage, Ia--CSM events can be detected out to tens of Mpc or, if absent, will constrain the majority of SNe Ia to $\dot{M}\lesssim 10^{-9}\,M_\odot\,{\rm yr^{-1}} $ for $v_w\sim100$\,km\,s$^{-1}$ (with the exact limits scaling with distance and microphysics). Figure~\ref{fig1} summarizes how these radio limits and detections map onto the single-degenerate mass-loss parameter space. The diagonal detectability boundaries reflect that the radio luminosity depends primarily on the wind density parameter $\dot{M}/v_w$: at higher wind velocities a proportionally higher mass-loss rate is required for the same CSM density, so higher $v_w$ does not make systems easier to detect.  VLBI confirmation will help distinguish genuine SN emission from host backgrounds and will measure the extremely small apparent sizes expected at early times, providing direct constraints on how efficiently the explosion energy couples to the CSM. A sizeable SKA+VLBI sample will allow us to map out the radio luminosity function of SNe Ia, quantify the contribution from different progenitor channels and test binary evolution models that lead to close double WD systems. This, in turn, feeds back into our understanding of the local distance scale and the role of SNe Ia as precision cosmological probes.

\subsection{Supernova remnants}
Supernova remnants (SNRs) remain detectable in radio for up to several thousand years after explosion and encode the time--integrated outcomes of SN explosions \citep{Ingallinera01.2026.SKA}. A complete SNR census plays a key role in measuring the delay time distributions of SNe and placing stringent constraints on the star formation history of their host galaxies. A few hundred SNRs have been detected in our Galaxy to date, far fewer than the several thousand predicted by population synthesis and Galactic SN-rate estimates \citep[e.g.,][]{2009BASI...37...45G}. 

This discrepancy likely reflects selection effects in past surveys. Limited angular resolution causes young, compact SNRs to be blended with bright backgrounds, while old, large remnants are intrinsically faint and often fall below the sensitivity limits of previous radio observations \citep{2017MNRAS.464.2326S}. SKA+VLBI naturally addresses both ends of this spectrum: SKA surveys will uncover faint, extended SNRs, while VLBI will resolve young, compact remnants, trace their morphology and proper motions, and measure post-shock magnetic structure on parsec and sub-parsec scales.   

For the closest young remnants, mas-resolution imaging will anchor the long time baseline between the first radio year and the thousandth, tying together SN, young SNR and mature SNR phases in a single, coherent evolutionary sequence. Comparing observed shell sizes, expansion rates and spectral indices with ejecta CSM/ISM interaction models will provide important clues to pre-SN mass-loss histories and explosion energetics, and will link the transient SN sky to the quasi steady radio sky of SNRs in a physically self-consistent way.

\section{Methodology \& Observational Strategy}
Our approach is to operate SKA--Mid as a commensal discovery and monitoring engine, and to promote a well defined subset of nearby, high value events to intensive SKA+VLBI follow up. 

\textbf{Discovery \& triggers.} 
Wide field SKA--Mid surveys at 1--5 GHz, revisiting large sky areas on day to week cadences, will either discover radio SNe directly or identify variable radio counterparts to optically discovered transients. For spectroscopically classified SNe within $\lesssim 50$ Mpc we will trigger rapid follow up ($\lesssim 24$ hr from discovery or classification) with SKA--Mid (continuum plus in-band spectra) to capture the rise through SSA, constrain the explosion epoch, and set a well-defined phase zero for subsequent kinematic modelling (see detailed discussion of rapid-response triggering for SKA radio transients in \citealt{GemmaAnderson01.2026.SKA}). These early epochs also establish whether the event is sufficiently radio bright and cleanly isolated from host emission to justify promotion to a full VLBI campaign.
Only events that meet brightness and field-quality criteria will be promoted to SKA+VLBI ToO observations, enabling a resource-efficient tiered strategy.

\textbf{VLBI imaging.}
We then implement multi-epoch sequences tailored to the expected shock speed, distance and subtype: at 10--30 Mpc, observations at $\approx$10, 30, 60, and 120 days at 1.6 GHz (sensitivity consideration) and 5--8 GHz (enhanced resolution, reduced scattering) jointly map both spectral evolution and the expanding ring/shell. For very nearby events ($\lesssim15$--20 Mpc), 8--15 GHz VLBI provides synthesized beams of order 0.6--0.9 mas on global baselines. Target selection prioritizes modest distance ($\lesssim 30$--50 Mpc), early time radio flux densities (tens to hundreds of $\mu$Jy), clearly non-thermal spectra, minimal host confusion, and maximum science payoff (stripped envelope and strongly interacting subtypes, plus rare SLSNe and Ia--CSM events). The phased SKA stations tied into EVN/VLBA/LBA increase baseline sensitivity by factor of several, making faint ($\lesssim 100~\mu$Jy) shells and compact cores feasible in $\sim$2--4 h tracks \citep{Chen01.2026.SKA, Kadler02.2026.SKA, Bempong-Manful01.2026.SKA}. The available VLBI beams support phase referencing and, when needed, in-beam calibration. Where suitable calibrators exist, multi phase center correlation will allow us to monitor serendipitous transients in the same field at negligible extra observing cost. Polarization is obtained whenever S/N allows, yielding constraints on post-shock magnetic fields via rotation measure synthesis. Data volumes of order terabytes per track are routine for modern correlators; established pipelines (AIPS, CASA, Difmap) combined with SKAO analysis tools and containerized, reproducible workflows will carry calibration and imaging through to publication quality products and legacy archives.

The program leverages multi-wavelength synergy. Optical and NIR surveys set explosion epochs and precise positions, while spectroscopy provides ejecta velocities and elemental diagnostics. X-ray facilities probe the thermal shock component and inverse Compton emission from the fastest ejecta. High-energy instruments test hadronic acceleration in dense CSM via gamma-ray signature. ALMA identifies dust and CO around interacting SNe and constrains the cold CSM reservoir.  Despite theoretical interest, GeV--TeV detections of young SNe remain elusive: \textit{Fermi}-LAT searches give upper limits on early emission, and H.E.S.S. likewise reports non-detections; CTA will transform the landscape, and radio-resolved shocks will be crucial to interpret any detections \citep{2015ApJ...807..169A,2019A&A...626A..57H}. SKA’s rapid response also prepares for the gravitational wave era of CCSN searches. If asymmetric core-collapse or fallback driven disk winds become detectable by third generation detectors, SKA+VLBI will provide the geometry resolving counterpart, tying any GW signal to a directly imaged shock and CSM configuration.

\begin{figure}
 \centering
\includegraphics[width=0.45\textwidth]{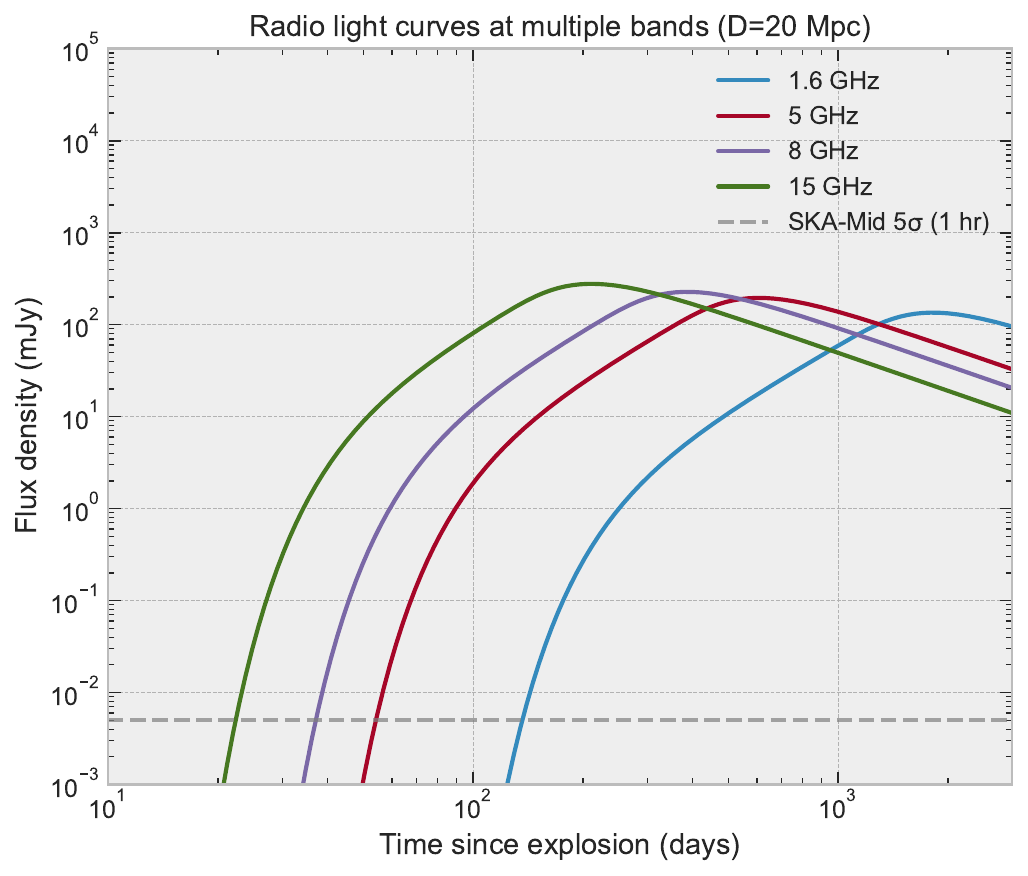}
\includegraphics[width=0.45\textwidth]{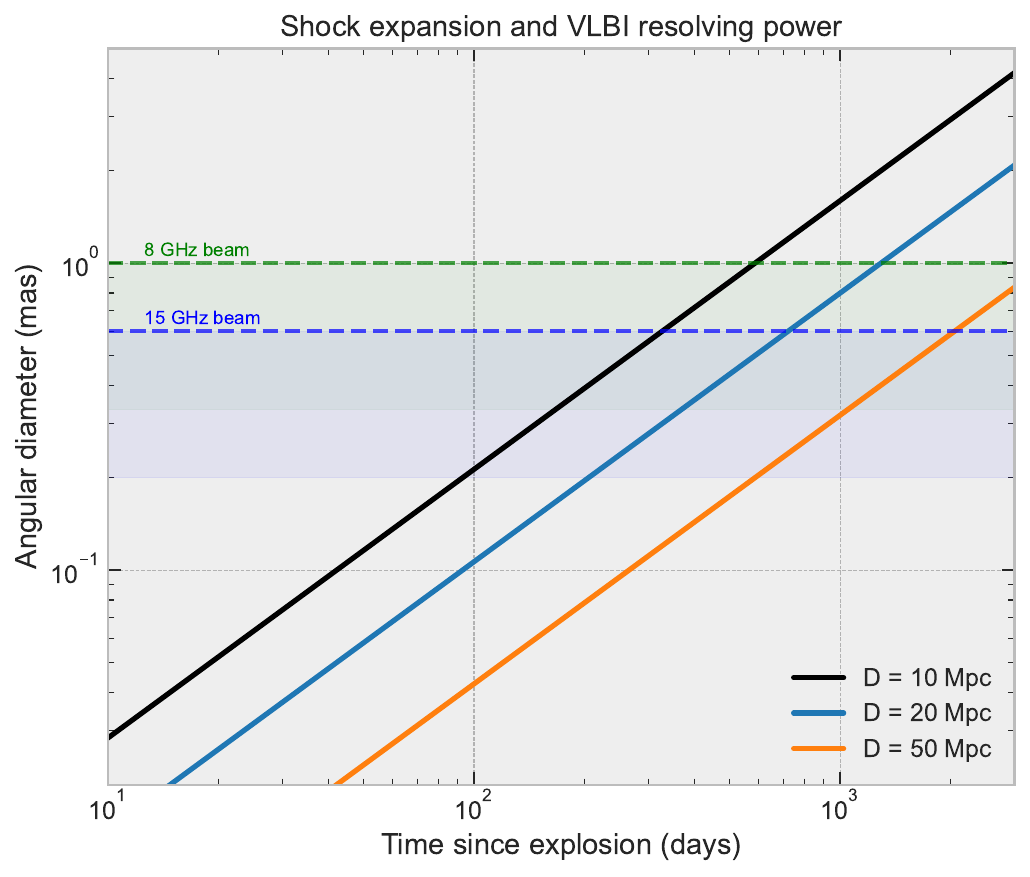}
\caption{
Left: Multi-frequency radio light curves of a core-collapse supernova at a fiducial distance of $D = 20$~Mpc, assuming a wind-like CSM with $\dot{M} = 10^{-4}\,M_\odot\,\mathrm{yr^{-1}}$ and $v_{\rm w} = 100\,\mathrm{km\,s^{-1}}$.
The 1.6, 5, 8, and 15~GHz curves include both synchrotron self-absorption (SSA) and external free--free absorption (FFA).
The horizontal dashed line marks a representative SKA-Mid $5\sigma$ detection limit for a 1~hr integration ($\sim 5~\mu$Jy), showing that typical events within $\sim 20$~Mpc can be followed over $10$--$300$~days across multiple bands, enabling constraints on the CSM density and shock evolution.
Right: Evolution of the model shock angular diameter as a function of time since explosion for $D = 10$, 20, and 50~Mpc.
Horizontal dashed lines indicate typical VLBI beam sizes at 8~GHz ($\sim 1$~mas) and 15~GHz ($\sim 0.6$~mas), with shaded regions marking the partially resolved regime ($\theta \gtrsim \mathrm{FWHM}/3$).
For nearby events with $D \lesssim 20$~Mpc, the shell becomes partially to fully resolved within a few tens of days, allowing direct measurements of the expansion law $R_{\rm sh} \propto t^{m}$, while at $D \sim 50$~Mpc resolving the ejecta--CSM interaction requires later epochs and/or more energetic explosions.
}

 \label{fig:sn_radio_LC}
 \end{figure}

Representative radio light curves and angular-size evolution for a fiducial
core-collapse SN model are shown in Figure~\ref{fig:sn_radio_LC}. For an event
at \(D = 20\) Mpc, SKA-Mid can follow the multi-frequency synchrotron emission
from the SSA/FFA-dominated rise through the optically thin decline over
\(\sim 10\)–300 days, while the angular diameter of the forward shock exceeds a
significant fraction of the VLBI beam (\(\theta \gtrsim {\rm FWHM}/3\)) within a few
tens of days for \(D \lesssim 20\) Mpc. This demonstrates that our proposed SKA+VLBI
strategy will routinely deliver both well-sampled light curves and partially
resolved shells for nearby SNe, enabling direct measurements of the expansion
law \(R_{\rm sh} \propto t^{m}\) and constraints on the progenitor mass-loss rate.

\section{Anticipated Breakthroughs}
(1) Kinematics with geometry. For a bright SN~IIb/Ib/Ic within 10--20 Mpc, four to five VLBI epochs at 5--8 GHz measure the angular radius and deceleration to $\approx 5$--$10\%$ and the axial ratio to $\approx 0.1$ precision at S/N$\gtrsim 10$, discriminating quasi-spherical from clearly bipolar solutions. Coupled to optical line velocities (which trace the bulk ejecta), these measurements break degeneracies between shock microphysics and CSM density profile, and calibrate the conversion from radio brightness temperature to physical energy partition between electrons, magnetic fields and kinetic energy.

(2) direct CSM imaging. In strongly interacting SNe within $\lesssim 50$ Mpc, VLBI will reveal limb-brightened shells of radius $\sim 0.2$--1.5 mas by 1--4 months and directly map the angular scale of pre-SN eruptions (LBV like shells) or dense equatorial zones created by binary interaction. The incidence and morphology of clumps, i.e., knotty fine structure with strong local brightening, provide the first statistical constraints on clump mass fractions and on angular patterns of mass loss. Comparing these maps with optical line profiles and X ray light curves will test whether the most extreme CSM configurations arise from brief eruptive episodes or from long lived, binary driven equatorial outflows.

(3) Compact cores. A handful of clean detections (flat-spectrum, compact, and with negligible expansion) would measure PWN birth rates and initial spin-down powers; a null result across many nearby IIP SNe would imply either long emergence times or relatively slow/weak pulsars at birth. In either case, SKA+VLBI turns individual detections or upper limits into quantitative constraints on the initial spin period distribution of neutron stars and on the fraction of CCSNe that immediately power observable central engines.

(4) SLSNe. SKA will determine whether magnetar‑powered SLSNe produce detectable radio nebulae after $\geq 1$ month; VLBI will discriminate central engines vs. CSM interaction if detected.   A sequence of deep 8--15 GHz epochs for a well defined SLSN sample within $D\lesssim200$ Mpc will either reveal compact, slowly expanding radio nebulae consistent with engine powered PWNe, or will push engine luminosities and baryon loading to levels that severely challenge standard magnetar models.

(5) SNe Ia. 
SKA+VLBI will measure Ia--CSM fraction (rare but clearly non zero, as exemplified by SN 2020eyj) and impose deep mass-loss limits (e.g., $\dot{M}\lesssim10^{-9}\,M_{\odot}\,{\rm yr^{-1}}$ for $100~{\rm km~s^{-1}}$ winds at $\lesssim30$--$50$ Mpc for non--detections).
Over a few years, we will know whether helium-rich donors are a percent-level minority or a more substantial channel. Depending on the pre-SN mass loss history, radio emission from SNe Ia could be intrinsically weak; the combination of high sensitivity and high angular resolution in SKA+VLBI will efficiently distinguish genuine Ia--CSM systems from host galaxy contamination and will sharply constrain the allowed parameter space of single-degenerate progenitor models. 

(6) SNRs. SKA will discover a large amount of `missing' SNRs in the Milky Way that were failed to be detected due to the limited-sensitivity and resolution of previous radio surveys. Given the large candidate set, our strategy is hierarchical: SKA surveys first identify SNR candidates via non-thermal spectra, polarization, and resolved shell-like morphology, and only the most compact and youngest remnants are then followed up with SKA+VLBI to measure expansion and proper motions. SKA+VLBI will give a tighter constraint on the star-formation rates of SNe host galaxies.  Resolved SKA+VLBI imaging of the youngest and most compact remnants will measure expansion velocities, shock morphologies and magnetic field structures, thereby turning the transient SN phase to the quasi steady SNR population and giving tighter, model based constraints on the star formation and core-collapse SN rates of nearby galaxies.

\section{Key Outcomes, Predictions, Legacy \& Impact}

SKA--Mid surveys at $\mu$Jy depths will yield of order a few hundred core--collapse radio SNe per year, with a few tens within $\lesssim 30$ Mpc suitable for routine SKA+VLBI imaging. In practice, global VLBI resources will favour a prioritized subset: most events will be monitored with SKA--Mid alone, while  the highest-value and best-placed targets (typically $\sim$10--15 per year) will be promoted to multi-epoch VLBI campaigns. This can be enabled in the SKA era through pre-allocated queue-scheduled ToO blocks, shorter 2--4 h tracks, and streamlined e-VLBI correlation and analysis. For this nearby sample, four 2--4 h epochs at 5--8 GHz will measure the expansion index $m$ and axis ratio $a/b$ to $5$–$10\%$, providing a quantitative test of type-dependent asymmetry, directly imaging CSM structure in IIn/Ibn events, and pushing most normal SNe Ia to mass–loss limits of $\dot{M}\lesssim10^{-9}\,M_{\odot}\,\mathrm{yr}^{-1}(v_{w}/100\,\mathrm{km\,s}^{-1})^{-1}$.
The same SKA+VLBI infrastructure will form a reusable time–domain engine for other explosive transients, while open, well–documented archives and common analysis workflows will train a new generation in time–domain VLBI and secure a long–lived legacy beyond the SKA era.


\begin{table}[t]
\centering
\caption{Mapping science goals to observing requirements. Sensitivities assume 1--4 hr per epoch with $t^{-1/2}$ scaling. Resolutions follow global baselines ($\approx 0.9$ mas@8 GHz; $\approx 1.5$ mas@5 GHz; $\approx 5$ mas@1.6 GHz).}
\footnotesize
\begin{tabular}{lcccc}
\hline\hline
Science goal & Distance & SKA--Mid role & SKA+VLBI role & Cadence \\
\hline
Geometry (IIb/Ib/Ic) & 10--25 Mpc & Early $\mu$Jy detection; spectra & 5--8 GHz imaging (1--2 mas) & 10--120 d \\
Strong CSM (IIn/Ibn) & 20--50 Mpc & Turnover \& SSA monitoring & 1.6+5 GHz shells/clumps & monthly, 2 yr \\
Compact cores (PWN) & $\leq 20$ Mpc & Late flattening & 8--15 GHz compactness & 100--300 d \\
SNe~Ia \& Ia-CSM & $\leq 50$ Mpc & Rapid, deep limits & Verify detections & 5--30 d \\
Luminosity source (SLSNe) & $\leq 200$ Mpc & Late-time radio emissions & 8--15 GHz & 100--200 d \\
\hline
\end{tabular}
\label{tab:reqs}
\end{table}

\section*{Acknowledgment} 
We thank the reviewer for the constructive comments. 
TA acknowledges the Shanghai Oriental Talent Project.
ZWL is supported by the CAS Project for Young Scientists in Basic Research (YSBR-148), the National Natural Science Foundation of China (NSFC, No. 12288102), the Strategic Priority Research Program of the Chinese Academy of Sciences (Nos. XDB1160303, XDB1160300, XDB1160000), the Yunnan Revitalization Talent Support Program ``YunLing Scholar" project and the International Centre of Supernovae (ICESUN), Yunnan Key Laboratory of Supernova Research (No. 202505AV340004).
MPT and JM acknowledge financial support from the Severo Ochoa grant CEX2021-001131-S and from the Spanish grant PID2023-147883NB-C21, funded by MCIU/AEI/ 10.13039/501100011033, as well as support through ERDF/EU. MPT and JMW also acknowledge the service and support from the Spanish Prototype of an SRC (SPSRC), funded by the Spanish Ministry of Science, Innovation and Universities, by the Regional Government of Andalusia, by the European Regional Development Funds and by the European Union NextGenerationEU/PRTR.
XCM is supported by the National Natural Science Foundation of China (12288102, 12333008), the Strategic Priority Research Program of the Chinese Academy of Sciences (XDB1160303, XDB1160000), and the National Science Foundation of China and National Key R\&D Program of China (2021YFA1600403), Yunnan Fundamental Research Projects (202401BC070007), International Centre of Supernovae, Yunnan Key Laboratory (202302AN360001), Yunnan Revitalization Talent Support Program -- Yunling Scholar Project and the Yunnan Revitalization Talent Support Program -- Science \& Technology Champion Project (202305AB350003), and the China Manned Space Program with grant No. CMS-CSST-2025-A13. 
AI tools were used solely to improve the language and readability of the manuscript.

\bibliographystyle{abbrvnat-maxbibnames4}
\bibliography{AASKAII_ID74_VLBI_SNe} 

\end{document}